\documentclass{article}
\usepackage{spconf,amsmath,graphicx}
\usepackage{amssymb}
\usepackage{amsfonts}
\usepackage{mathtools}
\usepackage{multirow}
\usepackage{bm}
\usepackage{nicefrac}
\usepackage{tabularx}
\usepackage{booktabs}
\usepackage{pifont}
\usepackage{amsthm}
\usepackage{extarrows}
\usepackage{hyperref}
\usepackage{balance}

\usepackage{array}

\def\s{{\mathbf s}}

\def\th{{\bm \theta}}
\def\sp{\hat{\mathbf s}}
\def\L{{\cal L}}

\def\R{{\mathbb R}}

\makeatletter
\def\hlinewd#1{%
  \noalign{\ifnum0=`}\fi\hrule \@height #1 \futurelet
   \reserved@a\@xhline}
\makeatother

\title{Unified Gradient Reweighting for Model Biasing \\ with Applications to Source Separation}
\name{Efthymios Tzinis${^{\natural, *}}$\thanks{*Equal contribution.
Code: \href{https://github.com/etzinis/biased_separation}{https://github.com/etzinis/biased\_separation} 
} \quad Dimitrios Bralios${^{\natural,\sharp,*}}$ \quad Paris Smaragdis$^{\natural,\flat}$}
\address{\hspace{-0.2cm}$^\natural$University of Illinois at Urbana-Champaign, $^\sharp$National Technical University of Athens,\ 
$^\flat$Adobe Research}
\begin{document}
\ninept
\maketitle
\begin{abstract}
Recent deep learning approaches have shown great improvement in audio source separation tasks. However, the vast majority of such work is focused on improving average separation performance, often neglecting to examine or control the distribution of the results. In this paper, we propose a simple, unified gradient reweighting scheme, with a lightweight modification to bias the learning process of a model and steer it towards a certain distribution of results. More specifically, we reweight the gradient updates of each batch, using a user-specified probability distribution. We apply this method to various source separation tasks, in order to shift the operating point of the models towards different objectives. We demonstrate different parameterizations of our unified reweighting scheme can be used towards addressing several real-world problems, such as unreliable separation estimates. Our framework enables the user to control a robustness trade-off between worst and average performance. Moreover, we experimentally show that our unified reweighting scheme can also be used in order to shift the focus of the model towards being more accurate for user-specified sound classes or even towards easier examples in order to enable faster convergence.

\end{abstract}
\section{Introduction}
\label{sec:intro}
One of the most fundamental problems in signal estimation is single-channel audio source separation, where one extracts the individual sources that constitute a mixture signal \cite{belouchrani1998blindsourceseparationTFrepresentations}. Recent deep learning approaches have greatly improved the state-of-the-art in supervised speech separation \cite{luo2019convTasNet, zeghidour2020wavesplit, liu2020causalCASA, nachmani2020fb_unknown_num_speakers}, music separation \cite{defossez2019music} as well as universal sound separation \cite{kavalerov2019universal, tzinis2020improving}. However, the reported performance is usually reported as the average signal-to-noise ratio (SNR) metric obtained between the estimated sources and the clean signals. In many real-world applications, users are mostly interested towards having a robust system where the failure cases are minimized or might prefer a biased system working best with specific sound classes.

The bias-variance trade-off has been studied for a long time \cite{geman199classic_bias_variance} and revisited for over-parameterized neural networks in \cite{belkin2019reconciling_bias_variance}. For instance, a fixed capacity separation model with low bias (performing well on average) might expect a larger variance when out of distribution sources appear (e.g. sources mixed at different SNRs or rare sound classes). To that end, an analysis of robust deep learning models with verifiable robustness guarantees has been conducted for image classification tasks \cite{zhang2019towards_verifiably_robust_nns} and bias correction terms have been introduced for weather estimation models \cite{moghim2017bias_correction_climate_model_temperature_precipitation}. Constructing more robust deep learning estimation models can also be achieved through an ensemble of models \cite{liu2018towards}. Specifically, a mixture of experts has been shown to yield improvements in terms of source reconstruction accuracy \cite{sivaraman2020mixture_of_experts_noob_paper}. In \cite{manilow2017predicting_SDR_for_ensembles}, a trainable selection mechanism which switches between biased separation models has shown to be effective under multiple separation tasks. Despite the great success in developing more reliable separation models, little progress has been made towards importing and exploiting biases for individual examples which are most important, for a given task, during training. 

Treating each sample equally during training might not be what one actually wants. Curriculum learning starts by using higher weights on ``easier" examples while gradually shifting to a uniform weight distribution and has been shown to significantly reduce training time as well as increasing generalization capabilities of models \cite{bengio2009curriculum_learning}. Reweighting of training examples has also been used for developing more accurate models by focusing on higher variance examples \cite{chang2017activebias_focus_onhigh_var_examples} and by using an adaptive meta-learning algorithm for noisy or imbalanced datasets \cite{ren2018learning_to_reweight_examples_for_robust_DL}. Importance sampling techniques on gradients have also been introduced in order to speed up the convergence of the optimization process \cite{katharopoulos18a_not_all_samples_equal} and achieve better generalization in semi-supervised tasks \cite{ren2020not_all_unlabeled_data_are_equal}. However, a recent study has shown that the effect of reweighting schemes for image classification networks might diminish when trained for a long time period and these schemes should only be employed in combination with an early stopping criterion \cite{byrd2019effect_of_importance_reweighting}. 

In this work, we treat the separation model under training as a partially fault tolerant system \cite{liu2018neural_fault_tolerant_adaptive} which is able to continue to operate well in certain cases while failing in others. Our premise is that for harder estimation problems such as source separation with multiple sound classes and limited computational resources, it might not be possible to achieve perfect reconstruction of all types of sources which might appear in an input mixture. Thus, biasing estimation models and shifting their operation point in order to make certain examples more significant than others is critical for real-world applications. We argue that the user should be able to handle this trade-off by seamlessly designing an algorithm that would best fit their application. For instance, a user might be mostly interested towards developing a more robust model, speeding up the convergence of their model, or ranking higher the importance of performing well on separating specific types of sources. The aforementioned previous work has tried to solve these types of problems only for detection tasks and by specifically designing ensembles or reweighting schemes for each case. We address these problems by following a more holistic approach and we make the following contributions:
\begin{enumerate}
    \item We formalize a unified gradient reweighting scheme that can be used by the users in order to shift the operation point of their estimation models for multiple applications.
    \item We present a simple way of defining a distribution over the gradients and show the efficacy of our method under a variety of separation tasks and operation modes.
\end{enumerate}

\section{Unified gradient importance reweighting scheme for biasing estimation models}

\subsection{Unbiased Estimation Model Training}
Any estimation model $f_{{\bm \theta}}$, with parameters ${\bm \theta}$, takes as input an observed signal $\mathbf{o} = \phi(\s) \in \R^{C \times T}$, where $C$ is the number of input channels, $T$ is the number of time samples and $\phi$ is a function provided by nature or artificially imposed that transforms the set of $N$ signals $\s \in \R^{N \times T}$ to the observed signal. Without loss of generality, we are going to describe the unified gradient reweighting approach for the case of source separation estimators. In this case, the observed signal is an input mixture of up to $N$ sounds $\mathbf{o} = \sum_{l=1}^{N} \s_l$ and we seek to train a separation model which estimates the individual sources as truthfully as possible $f_{{\bm \theta}} ( \mathbf{o} ) = \sp \in \R^{N \times T}$. For the cases where the mixture consists of less than $N$ sources, we assume that the inactive sources would be represented with zero vectors. Assuming the maximum number of sources in a mixture is a valid hypothesis also followed by the latest state-of-the-art approaches in source separation \cite{wisdom2020MixIT, nachmani2020fb_unknown_num_speakers}. In order to train this estimation model in a mini-batch stochastic gradient descend (SGD) sense, we minimize the loss $\L$ over a set of $B$ training examples in the batch, and perform the following update rule: 
\begin{equation}
\label{eq:unbiased_SGD}
    \begin{gathered}
    \th_{k+1} = \th_{k} - \eta \sum_{i=1}^{B} \frac{\mathbf{g}_k^{(i)}}{B}  , \enskip \mathbf{g}_k^{(i)} =  \nabla_{\th_{k}} \L \left( f_{{\bm \theta}} \left( \mathbf{o}^{(i)} \right), \s^{(i)} \right),
    \end{gathered}
\end{equation}
where $k$ denotes the optimization step, $\eta>0$ denotes the learning rate, and $( \mathbf{o}^{(i)} , \s^{(i)} ) \sim \mathcal{D}$ denotes the $i$th training sample in the batch drawn i.i.d. from the dataset $\mathcal{D}$. Usually the loss $\L$ measures the disparsity between the reconstructed sources $\sp$ and the clean sources $\s$. The aforementioned definition of optimization updates implicitly assumes that the gradients $\mathbf{g}_k^{(i)}$ are going to be equally weighted across all samples in the batch. Thus, the mini-batch noisy updates are going to be an unbiased estimator of the full gradient over the whole dataset. Therefore, the unbiased mini-batch updates for the $k$th optimization step, which are defined in Equation \ref{eq:unbiased_SGD}, are computed as the empirical mean of a discrete uniform distribution over the gradients of the samples in the batch:
\begin{equation}
\label{eq:unbiased_updates}
    \begin{gathered}
    \bm {\delta}_k = \underset{\mathcal{U} \{1,B\}}{\mathbb{E} } \left[ \mathbf{g}_k^{(i)} \right] = \sum_{i=1}^{B} \frac{1}{B} \mathbf{g}_k^{(i)}.
    \end{gathered}
\end{equation}

\subsection{Unified Gradient Reweighting for Biased Training}
Even though unbiased models have desirable properties, a user might want to bias the model towards specific examples that they are mostly interested in performing adequately. In this work, we present a unified biasing technique for multiple usages by reweighting the gradient updates of the examples in the batch. Specifically, we let the user specify a probability mass function $p$ over all samples in the batch. To this end, we are going to generalize the updates for the $k$th optimisation step, specified in Equation \ref{eq:unbiased_updates}, as shown next: 

\begin{equation}
\label{eq:general_biased_updates}
    \begin{gathered}
    \widetilde{\bm {\delta}}_k = \underset{p_k}{\mathbb{E} } \left[ \mathbf{g}_k^{(i)} \right] = \sum_{i=1}^{B} p_k\left(\mathbf{o}^{(i)}, \s^{(i)} \right)  \mathbf{g}_k^{(i)} \\ 
    \operatorname{s.t.} \enskip \sum_{i=1}^{B} p_k\left(\mathbf{o}^{(i)}, \s^{(i)} \right) = 1, \enskip \forall \enskip \left\{ ( \mathbf{o}^{(i)} , \s^{(i)} ) \right\}_{i=1}^B \in \mathcal{D},
    \end{gathered}
\end{equation}
where the probability mass function (pmf) $p_k$ can be any function that contains information regarding the training examples $( \mathbf{o}^{(i)} , \s^{(i)} ) \sim \mathcal{D}$ and can also change dynamically for successive optimization steps. Now the user is capable of controlling how the model would weight the contribution of each individual example based on any information that they consider vital.

\subsection{Softmax Gradient Reweighting}
\label{sec:method:SoftmaxReweighting}
Although the user is free to choose any arbitrary pmf in Equation \ref{eq:general_biased_updates}, we propose a simple but effective way of parameterizing this distribution using a softmax function. We showcase that by using a softmax parameterized family of distributions, one can instantiate different biasing configurations for their application. Here, we give a general form that we are going to use in this work:
\begin{equation}
\label{eq:general_softmax}
    \begin{gathered}
    p_k\left(\mathbf{o}^{(i)}, \s^{(i)} \right) = \frac{\operatorname{exp}(\operatorname{F}_k(\mathbf{o}^{(i)}, \s^{(i)} ))}{\sum_{j=1}^{B} \operatorname{exp}(\operatorname{F}_k(\mathbf{o}^{(j)}, \s^{(j)} )))}, \enskip \forall \enskip i, k, 
    \end{gathered}
\end{equation}
where $\operatorname{F}$ is a weighting function which will be instantiated according to the operation point that each user needs to shift the estimation model towards. Trivially, $p_k$ satisfies the constraints of Equation \ref{eq:general_biased_updates}.

\subsubsection{Robust Estimation}
In real-world systems, the expected performance is not the only metric that should be taken into consideration. The failure cases of an estimation model might be destructive for downstream tasks (e.g. a bad separation mechanism might lead to deficient performance of an ASR system). Although an estimator might reach its maximum capacity, we would like to shift its operating point towards reducing the amount of bad predictions. To that end, we follow a more system-wise approach by using our unified reweighting scheme to focus on the examples with higher loss value. Specifically, we introduce the following weighting function $\operatorname{F}$ as defined in Equation \ref{eq:general_biased_updates}:
\begin{equation}
\label{eq:bad_examples_weighting}
    \begin{gathered}
    \widehat{\operatorname{F}}_k(\mathbf{o}^{(i)}, \s^{(i)}) = \alpha \mathcal{L}(\sp, \s), \enskip \alpha > 0, \enskip  \forall \enskip i, k, 
    \end{gathered}
\end{equation}
where $\mathcal{L}$ is the signal-level separation loss and $\alpha$ is the robustness parameter, specified by the user, which is going to control how much the estimation model is going to be biased against producing totally erroneous estimations. It is trivial to show that by letting $\alpha \to 0$, the pmf becomes uniform and all gradients contribute equally. 

\subsubsection{Curriculum Training}
Sometimes the user is indifferent to the failure cases of their model and aims to bias the estimator initially towards ``easy" examples, that would help the model to converge faster. This technique falls under the category of \textit{curriculum learning} \cite{bengio2009curriculum_learning} and again can be expressed using our unified gradient reweighting scheme. Namely, we propose to use the following weighting function $\operatorname{F}$:
\begin{equation}
\label{eq:curriculum_weighting}
    \begin{gathered}
    \widetilde{\operatorname{F}}_k(\mathbf{o}^{(i)}, \s^{(i)}) = \beta(k) \mathcal{L}(\sp, \s), \enskip  \forall \enskip i, k, 
    \end{gathered}
\end{equation}
where $\beta(k)$ can be a function of the optimization step and controls the steepness of how the model is biased towards learning the ``easy" examples (with lower value of loss) first, and gradually converging to a uniform distribution.

\subsubsection{Bias Towards Specific Classes}
Another concern for estimation models is that in many cases the user is mostly interested in performing robust estimation towards specific classes while being very lenient towards failure cases on other classes. For instance, a real-world separation model on mobile devices might be used as a front-end for an ASR and an acoustic event detection system. Thus, reconstructing human voices might be more or less important given the specific application. In order to compensate for this kind of scenarios, we use our unified gradient reweighting scheme once more with an appropriate weighting function $\operatorname{F}$ that ranks the importance of different classes of sounds differently: 
\begin{equation}
\label{eq:class_weighting}
    \begin{gathered}
    \check{\operatorname{F}}_k(\mathbf{o}^{(i)}, \s^{(i)}) = \gamma(c^{(i)}), \enskip  \forall \enskip i, k, 
    \end{gathered}
\end{equation}
where $c^{(i)}$ denotes the class of sound of the $i$th example in the batch. As a result, the user might set values $\gamma(c^{(i)})$ that favor learning to reconstruct sounds from specific classes which are the most critical for a specific application while suppressing the contribution of others. This setup is extremely important for \textit{universal sound separation} tasks \cite{kavalerov2019universal} where multiple sound classes might exist in the training set.

\section{Experimental Framework}
\label{sec:Framework} 
To experimentally verify our approach we perform a diverse set of source separation experiments with multiple sound classes. 
\subsection{Audio Data and Mixture Generation Process}
We follow a similar training approach with \cite{tzinis2020twostep} and use two audio data collections in order to simulate real speech and environmental sound recordings. For the speech data collection we employ the widely used and publicly available WSJ0-2mix dataset \cite{hershey2016deepclustering} which consists of $14,823$ speech utterances from Wall Street Journal (WSJ0) corpus \cite{WSJ0}. In order to show the efficacy of our approach to arbitrary environmental sounds, we use the environmental sound classification (ESC50) data collection \cite{esc50} which consist of $50$ sound classes classes such as: \textit{non-speech human sounds}, \textit{natural soundscapes}, \textit{interior sounds}, \textit{animal sounds} and \textit{urban noises}. For each separation task, we generate our training mixtures using the augmented mixture generation process introduced in \cite{tzinis2020twostep}. This online augmentation process enables the generation of much more diverse mixtures by randomly choosing sounds from distinct audio classes, cropping of $4$ seconds segments and mixing sources at uniformly random SNRs in $[-30, 30]\, \mathrm{dB}$. For each epoch, we generate $20,000$ new mixtures while there are in total $5,000$ and $3,000$ mixtures for validation and testing, correspondingly. All audio files are downsampled at $8$kHz.

\subsection{Separation Network Architecture}
\label{sec:Framework:sep_network}
We use a mask-based time-dilated convolutional architecture as a separation network. Specifically, we use the \textit{Sudo rm -rf} architecture \cite{tzinis2020sudo} which has been shown to perform comparably with other state-of-the-art \cite{luo2019convTasNet, nachmani2020fb_unknown_num_speakers} architectures with a much lower computational complexity in terms of memory, number of parameters and number of floating point operations for both speech and environmental sound source separation tasks. For the encoder and decoder modules we use a kernel with a length of $41$ samples and stride of $20$ samples, corresponding to $2.625$ms and $1.28$ms, respectively. We set the number of encoder and decoder basis equal to $512$ and the number of U-ConvBlocks to $8$ where in each one there are $4$ successive downsampling and upsampling operations. The values were set empirically in order to obtain a good trade-off between performance and efficiency. It is important to underline that our unified gradient reweighting approach is not dependent on the estimation model that we use and can be applied effortlessly to any other architecture.

\subsection{Training and Evaluation Details}
\label{sec:Framework:train_eval_details}
Permutation invariant scale-invariant signal-to-distortion-ratio (SI-SDR) \cite{le2019sdr} measures the fidelity of the reconstructed sources $\sp$ w.r.t. the clean sources $\s$. Formally, we define this signal-level metric as:
\begin{equation}
\label{eq:SISDRloss}
    \begin{gathered}
    \text{SI-SDR}(\sp, \s^*) = - 10 \log_{10} \left(\| \rho \s^*\|^2 / \| \rho \s^* - \sp\|^2 \right),
    \end{gathered}
\end{equation}
where $\textbf{s}^*$ denotes the permutation of the sources that maximizes the mean SI-SDR across sources and $\rho =  \hat{\textbf{s}}^\top  \textbf{s}^* /\|\textbf{s}\|^2$ is just a scalar. For evaluating the performance of each model under a separation task, we report the SI-SDR improvement (SI-SDRi) which is the gain that we get on the SI-SDR measure using the estimated signals $\sp$ instead of the input mixture signal (observed signal $\mathbf{o}$). During training, the signal level loss function which is minimized is the negative SI-SDRi $\mathcal{L} = -\text{SI-SDR}(\sp, \s^*) + \sum_{j=1}^N \text{SI-SDR}(\mathbf{o}, \s_j)$. For each example in the batch, we are using the value of this loss for reweighting the gradients (see Section \ref{sec:method:SoftmaxReweighting}). 

From Section \ref{sec:method:SoftmaxReweighting}, it is evident that our approach heavily relies on having a batch size which can appropriately fit many examples which are drawn i.i.d. from the dataset. For that reason, we maximize the number of examples in the batch by setting the batch size equal to $B=28$ in order to be able to train each model on a single Nvidia GeForce RTX 2080 Ti graphics processing unit (GPU). In order to show that our gradient reweighting approach is also compatible with state-of-the-art optimizers, we use the Adam optimizer \cite{adam} with an initial learning rate set to $0.001$ and train our models for $100$ epochs. We also perform gradient clipping \cite{seetharaman2020autoclip} when the norm exceeds $5$.

\subsection{Source Separation Modes of Interest}
\label{sec:Framework:ModesOfInterest}
We test our unified gradient reweighting scheme for the cases presented in Section \ref{sec:method:SoftmaxReweighting} which are potential modes of interest for biasing estimation models. For each case below, we analyze the configuration of the weighting function $\operatorname{F}$ and the separation task where we assume that two sources exist in each mixture.

\subsubsection{Robust Estimation}
\label{sec:Framework:ModesOfInterest:Robust}
In this experiment, we use our reweighting scheme to improve the robustness of our separation networks and show the effectiveness of our approach under environmental sound separation. We try different values of the parameter $\alpha \in [0, 0.2]$ (see Equation \ref{eq:bad_examples_weighting}) in order to demonstrate how we can adjust the robustness of our models.

\subsubsection{Curriculum Training}
\label{sec:Framework:ModesOfInterest:Curriculum}
We extend the usage of our reweighting scheme by showing how one can achieve faster convergence in a speech separation task. The network is trained using the weighting function $\widetilde{\operatorname{F}}$ as defined in Equation \ref{eq:curriculum_weighting}, where the parameter is defined as  $\beta(\kappa) = -  \left(10 + 0.5 \kappa \right)^{-1}$ and $\kappa$ is the epoch count index. This is equivalent to a ``curriculum learning" \cite{bengio2009curriculum_learning} setup where greater weight is initially assigned to the examples that achieve a high SI-SDRi, and gradually the weight distribution becomes uniform. We compare this scheme, which could be considered curriculum training, to the baseline ($\beta(\kappa) = 0$).

\subsubsection{Bias Towards Specific Classes}

In the final experiment, we train separation models using mixtures that contain one speech utterance and one environmental sound. In order to bias the model towards a specific class $c$, we change the corresponding class parameter $\gamma(c)$ in the weighting function $\check{\operatorname{F}}$ (see Equation \ref{eq:class_weighting}). We compare a speech oriented model and a environmental sound oriented model against the unbiased model.

\label{sec:Framework:ModesOfInterest:Classes}

\section{Results \& Discussion}

\subsection{Robust Sound Separation}
\begin{figure}[!h]
    \centering
      \includegraphics[width=\linewidth]{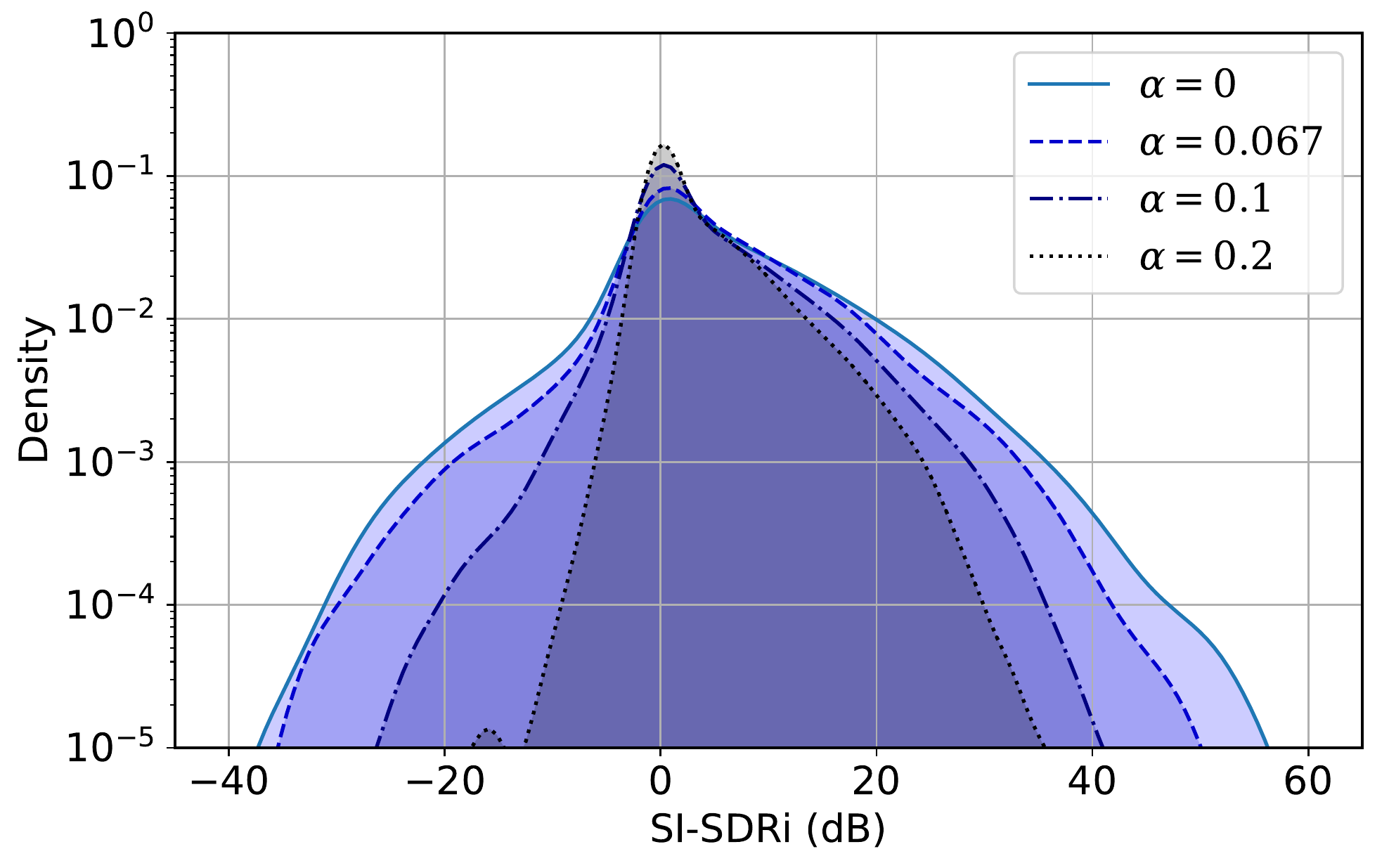}
      \caption{Distribution of test SI-SDRi (dB) in log-scale for environmental sound separation using different values of $\alpha$ in the weighting function $\widehat{\operatorname{F}}$. Note that by increasing $\alpha$, we put a higher weight to examples with higher loss. Essentially, we control the trade-off space between robustness and mean accuracy of the separation model.}
      \label{fig:robust}
\end{figure}

As previously specified, in this mode we take into consideration the failure case of separation systems. Specifically, we explore the trade-off between performing well on average and making a system more robust. In Figure \ref{fig:robust}, the distribution of the SI-SDRi of all estimations is depicted. It is evident that by changing the parameter $\alpha$ (see Equation \ref{eq:bad_examples_weighting}), we make the system more robust towards bad examples by sacrificing the performance on easier examples. The latter claim is also reflected in Table \ref{tab:robust_separation} where we also report the SI-SDRi performance in terms of multiple quantiles. For instance, the worst $1\%$ of examples in the test set, score an SI-SDRi less than $-17.5$dB when using a uniform distribution over samples. However, when we penalize the bad examples by using our gradient reweighting scheme with $\alpha=0.2$, the performance of the same quantile is increased to $-2.8$dB. We want to underline that this aspect of separation models is often neglected, but in many real-world applications a more robust model might be preferred over a more accurate (on average) model with higher variance. Depending on the application and the desired level of robustness, the user might shift the parameter $\alpha$ in our gradient reweighting scheme in order to shift the operation point of their separation system. 

\begin{table}[!htb]
    \setlength{\tabcolsep}{2.825pt}
    \centering
    \begin{tabular}{l|cc|ccccccccc}
    \toprule
     \multicolumn{12}{c}{Test SI-SDRi (dB)} \\    \hlinewd{1pt}
    \multirow{2}{*}{$\alpha$} & \multicolumn{2}{c|}{Statistics} & \multicolumn{9}{c}{Quantiles} \\\cline{2-12} 
     & mean & std & 1 & 5 & 10 & 25 & 50 & 75 & 90 & 95 & 99 \\
    \hlinewd{1pt}
    0 & \textbf{5.0} & 9.0 & -17.5 & -6.9 & -2.5 & 0.0 & \textbf{2.7} & \textbf{9.5} & \textbf{17.1} & \textbf{22.0} & \textbf{31.5} \\
    $\nicefrac{1}{15}$ & 4.6 & 7.8 & -14.7 & -4.4 & -1.4 & \textbf{0.1} & 2.4 & 8.4 & 15.2 & 19.4 & 29.0 \\
    $\nicefrac{1}{10}$ & 3.6 & 6.0 & -6.5 & -2.2 & -1.0 & -0.1 & 1.1 & 6.1 & 12.1 & 16.1 & 23.5 \\
    $\nicefrac{1}{5}$ & 3.0 & \textbf{4.8} & \textbf{-2.8} & \textbf{-0.9} & \textbf{-0.3} & 0.0 & 0.6 & 4.9 & 9.8 & 13.2 & 19.8 \\
    \bottomrule
    \end{tabular}
     \caption{Environmental sound separation performance metrics averaged over ten epochs while varying the robustness parameter $\alpha$.}
    \label{tab:robust_separation}
\end{table}

\subsection{Curriculum Training}
\begin{figure}[!htb]
    \centering
      \includegraphics[width=\linewidth]{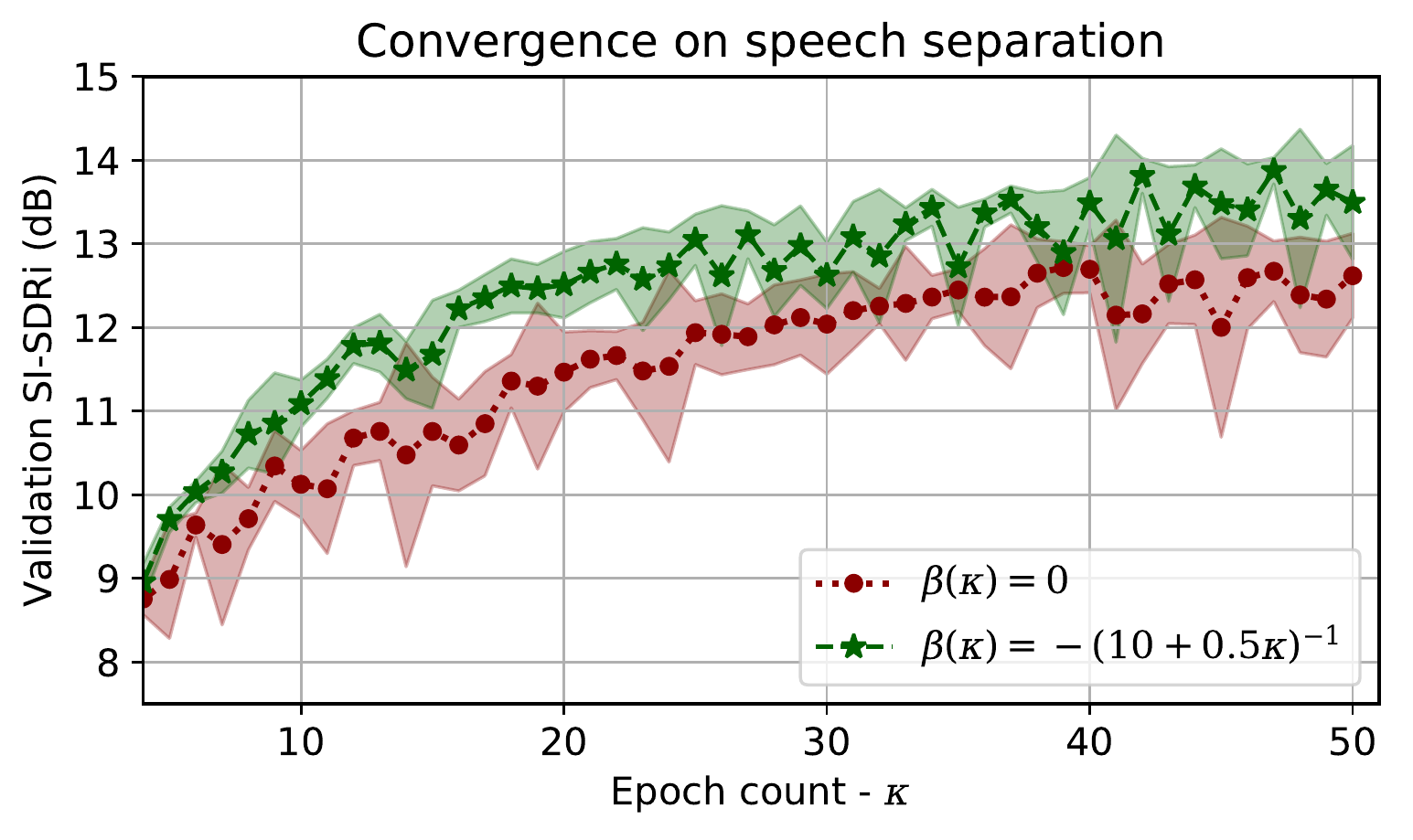}
      \caption{Mean Validation SI-SDRi (dB) across epochs computed over $5$ runs with random initializations. The shaded region denotes two standard deviations from the mean for those observations.}
      \label{fig:convergence}
\end{figure}

In Figure \ref{fig:convergence}, we show the average validation SI-SDRi obtained for two configurations over time. Notably, our unified gradient reweighting scheme can also be used in order to perform ``curriculum training" by setting the appropriate weighting function $\widetilde{\operatorname{F}}$. The gradient reweighted configuration yields a much faster convergence for the same number of training epochs compared to the baseline.

\subsection{Biasing Towards Specific Classes}
The separation performance results are presented in Table \ref{tab:mixed_separation}. We observe that the environmental sound oriented model outperforms the other two on examples coming from the same class. Similarly, the speech oriented model achieves a better performance on speech estimations, as well as overall. We hypothesize that the latter behavior is caused by the ``easiness" of speech examples that cause the model to understand faster how to aptly separate sources.  

\begin{table}[!htb]
    \centering
    \begin{tabular}{cc|ccc}
    \toprule
    \multicolumn{2}{c|}{$\gamma$} & \multicolumn{3}{c}{Mean test SI-SDRi (dB)} \\ \cline{3-5}
    Speech & Env. & Speech & Env. & Combined \\
    \hlinewd{1pt}
    0 & 0 & $12.2 \pm 0.1$ & $13.1 \pm 0.1$ &  $12.7 \pm 0.1$ \\
    0 & 3 &  $11.8 \pm 0.2$ &  $\mathbf{13.5 \pm 0.1}$ &  $12.6 \pm 0.1$ \\
    3 & 0 & $\mathbf{12.7 \pm 0.1} $ &  $13.1 \pm 0.1$ & $\mathbf{12.9 \pm 0.1} $ \\
    \bottomrule
    \end{tabular}
    \caption{Separation performance on individual sound classes and combined while training with different weights $\gamma(c)$ for \textit{Speech} and \textit{Environmental Sounds} (Env.). The mean is computed over the 5 last epochs while also showing one standard deviation from the mean.}
    \label{tab:mixed_separation}
\end{table}

\section{Conclusion}
We have presented a simple and unified gradient reweighting scheme with negligible computational cost that enables users and practitioners to shift the operation point of separation models. We have experimentally showed that we can use our proposed reweighting strategy in order to adjust the robustness of estimation models, train faster by focusing on easier examples and bias models towards specific classes under multiple separation tasks. Our approach remains general enough in order to be applied towards many detection, estimation or generation problems with minimal modifications.

\bibliographystyle{IEEEbib}
\bibliography{refs}
\end{document}